# Comment on "Ghost imaging with a single detector" [arXiv0812.2633v2]

Morton H. Rubin

*Department of Physics, University of Maryland, Baltimore County, Baltimore, Maryland 21250, USA*

Computational ghost imaging has been demonstrated experimentally recently. In this comment we wish to clarify the difference between pseudothermal ghost imaging (PGI) and computational ghost imaging (CGI). In particular, to emphasize the physics that arises from the difference in the type of sources used. The experiment in [2] is not a PGI experiment but is a simulation of PGI. If we consider CGI at the single photon level, it becomes clear that CGI relies on two identical photons, one real and one simulated, to obtain an image.

Computational ghost imaging (CGI) was introduced in [1] and experimentally demonstrated in [2]. The term ghost imaging originally arose in connection with quantum ghost imaging (QGI) to indicate that an image can emerge from the correlation between the output of a bucket detector that collects light that interacted with an object with the output from a scanning point detector or a CCD array that is illuminated by the same source [3]. The bucket detector has no imaging capability. In CGI the output of the second detector is replaced by a set of computational outputs. Pseudothermal ghost imaging (PGI) uses a single random source and, by correlating the output of two detectors, obtains an image of an object [4,5,6]. The key distinction between CGI and PGI lies in the sources. CGI uses a special set of incoherent sources (we shall refer to this as the C-source) while PSI uses a single pseudothermal random source. For each realization of the CGI measurement more information about the field generated by the source must be known beyond the information required for PSI. This information ensures that each real photon detected is correlated with an identical simulated photon. The statistics of PGI processes can be computed from the statistics of a classical source, that is, a source with a density matrix that can be described by a positive Glauber-Sudarshan P-function, however, PGI is at a fundamental level a two-photon interference phenomenon. We want to stress that this is more than a statement that all physics is fundamentally quantum mechanical, rather it is a statement about the underlying physics of PGI and leads us to conclude that CGI is simulation of PGI. To emphasize what we can learn from understanding these processes at a fundamental level, recall that the Young double slit experiment is completely explained classically; however, the fact that it can be performed one photon at a time leads to the important physics, as Dirac pointed out a long time ago, that each of the photons in the experiment interferes with itself. In the same way, the original QGI experiment [3] shows that a biphoton interferes with itself. PGI can be carried out with a source at sufficiently low intensity so that only two photons at a time are present in the system. Clearly, a minimum of two photons is needed to obtain a coincidence between the detectors. Repeating the experiment many times will eventually generate an image. We shall see below how CGI mimics this.

If the PSI source is replaced with a C-source of the type envisioned in [1], it is clear that this experiment can also be done two photons at a time, as demonstrated in [1]; however, as also demonstrated in [1], it can also be done using one real photon at a time. In order to clarify the difference between PGI and CGI let us examine how the image is formed. For details about the conditions under which the experiment is performed see [6]. Let $G_{xj}$ be the propagator from a source point $j$ to the point $x$ on the object. The object plane will be taken to be the surface of the bucket detector. Let $g_{yj}$ be the propagator from a point $j$ on the source to a point detector on a CCD array. It has been shown many times, in particular in the low intensity limit [6], that the intensity-intensity correlation may be written as

$$\langle I_b I_y \rangle = \sum_{j,k,x} \left| G_{xj} g_{yk} + G_{xk} g_{yj} \right|^2$$

$$= \langle I_b \rangle \langle I_y \rangle + \left| \sum_{j,x} G_{xj} g_{yj}^* \right|^2 \quad (1)$$

where $\langle I_b \rangle = \sum_{j,x} |G_{xj}|^2$ and $\langle I_y \rangle = \sum_{j} |g_{yj}|^2$. Now consider the Fig. 1. The source S illuminates the area AB on the beamsplitter BS. Now consider a photon that impinges on BS, there is a probability that it will be detected at $y$ by a point detector on the plane of the CCD detector and an equal probability that it will be detected at the point $x$ on the object plane. For simplicity we consider a one-dimensional system. It is clear that with a single photon we cannot detect any interference that can generate an image between these two probability amplitudes[†]. However, if a second, identical photon is emitted within the coherence time of the source, it is possible to see the interference. This is what the second term in the second line of eq. (1) is telling us. It is useful to use the Klyshko picture to understand this. Let the bucket detector BD be an incoherent plane wave source illuminating the object. Light transmitted through the point x on the object propagated backwards in time to the point j on the source by the propagator $G^*_{xj}$ it then is propagated forward in time to the point y by $g_{yj}$ and the total process $x \rightarrow y$ is given by $\sum_j G^*_{xj} g_{yj}$. The key to the imaging property is that the point x is uniquely determined by y in an ideal system. For a real system, there is a point spread function determined by diffraction so the mapping, $x \rightarrow y$, is really a point to spot mapping. The point spread function is determined by diffraction. Therefore, even though a bucket detector is used behind the object, the imaging mapping is not destroyed. In short, the detection of a photon at $y$ is correlated with the detection of an identical photon at $x$. It has been noted that in the Klyshko picture the source acts like a phase conjugate mirror [6,7].

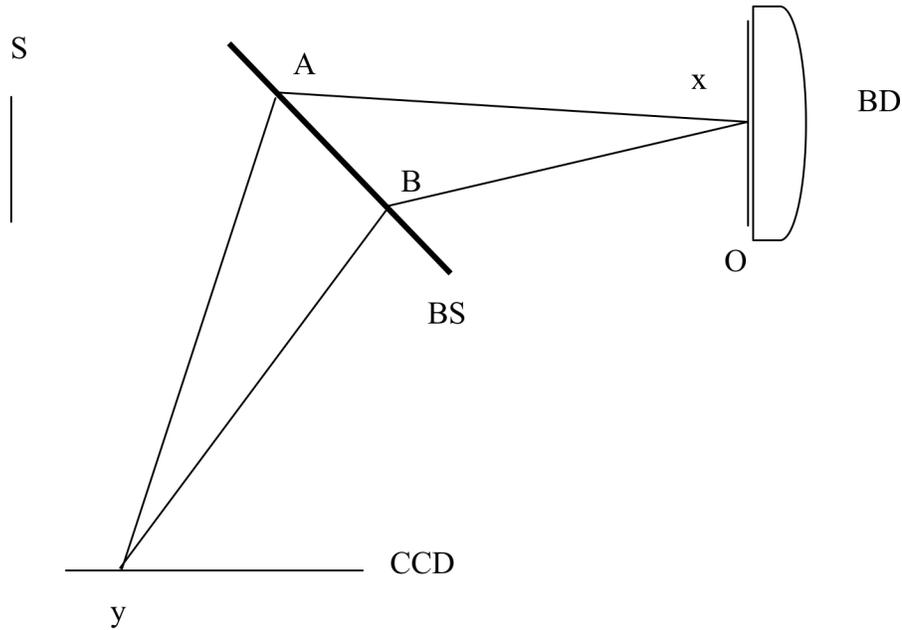

Figure 1: S is a pseudorandom source that illuminates the 50-50 beamsplitter BS, y is a point detector on the CCD detector that is located at a distance from the BS equal to the distance to the object. The object O lies on the face of the bucket detector BD.

Let us now see how CGI is related to PGI using the picture just developed. It is not difficult to see that a photon that is detected by the bucket detector that was transmitted through the object at $x$ was generated by a specific field distribution on BS. That field can be propagated back to a particular field distribution on the source. Each pair of (post-selected) points $x$ and $y$ corresponds to a distinct such distribution. Therefore, so long as we know what the distribution is, we can compute the intensity at $y$, multiply it $\int dx |t(x)|^2 I_x$ where $t(x)$ is the object's transmittance and repeat this many times. Then as stated in [2], we get $\langle I_x I_y \rangle \sim I_x^2 \delta(x-y)$. This gives the image $|t(x)|^2$. The specific set of source distributions is what composes the C-source. For each measurement, it is necessary to know the particular source used. Now suppose that the experiment [2] were carried out at the single photon level. If the C-source intenity is reduced to mimic a single photon source, then an image would still emerge because, by knowing which particular distribution was used in the measurement, a field due to an identical photon can be computed. Using the same source distribution is how CGI ensures that the interference is between two identical photons, one real and one simulated.

In conclusion, CGI may be thought of as a type of simulation of PGI. Both processes require interference between two photons to generate the image. A careful analysis of the physics of CGI and PGI shows that the statement in [2] at the end of the second paragraph that their measurements demonstrates that PGI and PGD "cannot possibly depend on any non-local quantum correlations" must be understood in terms of the definition of quantum correlations. If the authors mean the statistics can be computed using a positive P-function, we agree; however, if they mean it is not a two photon interference phenomenon, then we disagree.

The author thanks Yanhua Shih and Jianming Wen for helpful discussions about this comment. This work was support in part by U.S. ARO MURI Grant W911NF-05-1-0197.

References

† It might be thought that one could replace the detector at *y* by a pinhole and recombine the beams to do Young double slit type experiment. However, recall that there is no slit at the bucket detector and no image formation is possible.